\documentclass[prb,twocolumn,preprintnumbers,amsmath,amssymb,superscriptaddress]{revtex4}
\usepackage{graphicx}
\usepackage{latexsym}
\usepackage{amsmath}
\usepackage{graphics}
\usepackage{amssymb}
\usepackage{layout}
\usepackage{verbatim}
\usepackage{amsfonts,epsfig}
\usepackage{color}
\newcommand{\ket}[1]{\left|#1\right>}
\newcommand{\bra}[1]{\left< #1 \right|}
\newcommand{\beq}{\begin{equation}}
\newcommand{\eeq}{\end{equation}}
\newcommand{\bea}{\begin{eqnarray}}
\newcommand{\eea}{\end{eqnarray}}

\newcommand{\PbSnTe}{Pb$_{1-x}$Sn$_{x}$Te{}~}

\begin{document}

\title{PbTe/PbSnTe heterostructures as analogs of topological insulators}
\author{Ryszard Buczko}
\affiliation{Institute of Physics, Polish Academy of Sciences, al.~Lotnik{\'o}w 32/46, PL 02-668 Warszawa, Poland}
\author{{\L}ukasz Cywi{\'n}ski}
\affiliation{Institute of Physics, Polish Academy of Sciences, al.~Lotnik{\'o}w 32/46, PL 02-668 Warszawa, Poland}
\date{\today }

\begin{abstract}
We investigate theoretically the PbTe/\PbSnTe heterostructure grown in [111] direction, specifically a quantum wall (potential step of width $d$) of PbTe embedded in Pb$_{1-x}$Sn$_{x}$Te. For $x$ large enough to lead to band inversion, and for large $d$, there are well-known gapless interface states associated with four $L$ valleys. We show that for $d\!\approx\! 10$ nm the three pairs of states from oblique valleys strongly couple, and become gapped with a gap $\sim\! 10$ meV. On the other hand, the interface states from the [111] valley are essentially uncoupled, and they retain their helical character, remaining analogous to states at surfaces of thin layers of three-dimensional topological insulators. This opens up a possibility of studying the physics of two-dimensional helical Dirac fermions in heterostuctures of already widely studied IV-VI semicondcutors.
\end{abstract}

\maketitle

\section{Introduction.} 
A large attention has been recently devoted to topological insulators (TIs), materials in which the topology of the bulk bandstructure guarantees the existence of robust (against non-magnetic disorder) spin-nondegenerate (helical) edge/surface states.\cite{Hasan_RMP10,Qi_RMP11} TIs are essentially a subclass of narrow gap semiconductors, since their bandgaps must be smaller than typical energy scale of relativistic corrections to the bandstructure ($< \! 1$ eV). In fact, the first two-dimensional TI was realized\cite{Konig_Science07} in a quantum well of HgTe/HgCdTe, i.e.~a heterostructure based on well-known compound narrow-gap semiconductors. Helical edge modes have also been recently reported in a heterostructure of InAs/GaSb.\cite{Knez_PRL11}

Part of interest in TIs stems from the fact that they represent a new topological phase of noninteracting electrons: the TI character of a material is its bulk property, nontrivially encoded in the wavefunctions of the occupied (valence band) states. However, it is the presence of the helical edge/surface states which leads to observable consequences. In three dimensional strong (weak) TIs  they consist of an odd (even) number of spin-nondegenerate Dirac cones,\cite{Fu_PRB07} and many interesting effects were predicted to occur when these states  become gapped due to a perturbation,\cite{Hasan_RMP10,Qi_RMP11} e.g.~due to proximity to a superconductor \cite{Fu_PRL08} or a ferromagnet.\cite{Qi_PRB08,Essin_PRL09} 
The superconducting proximity effect was predicted to lead to a creation of Majorana excitations at the surface of a TI,\cite{Fu_PRL08} while covering the surface with a magnetic insulator was predicted to lead to new magnetoelectric effects.\cite{Qi_PRB08,Essin_PRL09}
By now many materials have been shown to be 3D TIs (see Ref.~\onlinecite{Hasan_RMP10} and references therein), with Bi$_{2}$Se$_{3}$ \cite{Xia_NP09} and Bi$_{2}$Te$_{3}$ \cite{Chen_Science09} gathering the most attention due to the feature of having a single Dirac cone at the surface. 
Despite a recent progress in gating of thin layers of Bi$_{2}$Se$_{3}$, \cite{Checkelsky_PRL11,Sacepe_NC11,Kim_NP12} it would be desirable to investigate the helical surface states in a well-known semiconductor system, for which the growth and nanostructure processing are already mastered. One such platform is strained bulk HgTe.\cite{Brune_PRL11} In this article we propose to focus on heterostuctures of compound semiconductors from  the IV-VI lead chalcogenide family, (Pb,Sn)Te.\cite{Nimtz}

The superlattices of PbTe/\PbSnTe (with $x \! \leq \! 0.18$) were thoroughly characterized.\cite{Kriechbaum_PRB84} Due to the huge dielectric constant of PbTe, mesoscopic structures of PbTe/PbEuTe exhibit very robust ballistic transport properties.\cite{Grabecki_PRB05,Grabecki_PE06,Kolwas_arXiv11} 
Both Mn and Eu can be incorporated into PbTe, and exchange coupling between their spins and the band carriers was studied.\cite{Dietl_PRB94} Furthermore, there exist heterostructures of PbTe and europium chalcogenides\cite{Kepa_PRB03,Lechner_PRL05} such as EuSe and EuTe, which are magnetic semiconductors with rich phase diagrams. Finally, a superconducting proximity effect has been recently shown to exist at In-PbTe junctions.\cite{Grabecki_JAP10} All these features suggest that IV-VI based heterostructure, if it hosted gapless helical states, could be a good candidate for investigations of phenomena related to TIs.

When changing the Sn content in \PbSnTe~alloy, band inversion between the topmost valence band and the lowest conduction band occurs at $x\! \approx \! 0.37$. In an unstrained sample this happens simultaneously at the four $L$ points at the edges of the Brillouin zone. Since the topological class of the bandstructure changes at the instances of local bandgap closing,\cite{Murakami_PRB08} after an even number of local band inversions this class remains the same. In fact, both PbTe and PbSnTe are in a topologically trivial class,\cite{Fu_PRB07} although very recently it was argued\cite{Hsieh_arXiv12} that SnTe is a topological crystalline insulator,\cite{Fu_PRL11} i.e.~that it supports an even number of surface states for specific surface orientations.
However, we focus here on the \emph{mutual}  band inversion between PbTe and PbSnTe, which leads to existence of interfacial helical states, predicted theoretically 25 years ago.\cite{Volkov_JETP85,Korenman_PRB87,Agassi_PRB88,Pankratov_SST90}

Such an interface of PbTe and \PbSnTe, at which four Dirac cones appear, is analogous to a surface of a weak TI. Recent research on weak TIs suggests that these cones are in fact protected against time-reversal invariant perturbations as long as these are not periodic and commensurate with the lattice spacing,\cite{Ringel_arXiv11,Mong_PRL12} 
a situation which is rather improbable at an epitaxial interface of two materials with the same lattice structure. It would however be even more interesting if one could obtain a single gapless helical state at an interface. As we show in this article, this can be brought by a confinement effect in a properly designed PbTe/\PbSnTe heterostructure grown in [111] direction. This happens due to an anisotropic energy structure near the four L points, which leads to a much stronger coupling of oblique valley states from opposite interfaces, compared to the [111] valley along the growth direction. Note that a similar coupling opens up a gap in a thin layer of a TI.\cite{Liu_crossover_PRB10,Linder_PRB09,Lu_PRB10,Shan_NJP10} With a proper choice of strain, composition, and layer width, we obtain practically gapless helical states for [111] valley, with Dirac points located in the finite-width-induced gap in the remaining three valleys.

The paper is organized in the following way. In Section \ref{sec:H} we describe the $\mathbf{k}\! \cdot \! \mathbf{p}$ Hamiltonian at L-point band extrema in PbSnTe, and we note its close relation to the $\mathbf{k}\! \cdot \! \mathbf{p}$ Hamiltonian describing the states at the $\Gamma$ point in Bi$_{2}$Se$_{3}$-type materials. In Section \ref{sec:helical} we review the properties of helical states existing at the PbTe/\PbSnTe interface, and we re-express some known results\cite{Volkov_JETP85,Pankratov_SST90,Korenman_PRB87,Agassi_PRB88} in a manner that shows their complete analogy with the surface states of TIs. Then, in Section \ref{sec:width} we present the main results of the paper,  the calculation of electronic structure of a finite-width PbTe/\PbSnTe heterostructure. There we present our prediction that in a properly designed heterostructure one can have  a practically gapless single Dirac cone, with its Dirac point located in the gaps of the states from the other valleys. Finally, in Section \ref{sec:discussion} we discuss a few practical issues related to the experimental realization of the proposed heterostructure.

\section{The Hamiltonian.} \label{sec:H}
The electronic states near the L-point extrema in \PbSnTe are described by Dimmock's Hamiltonian.\cite{Nimtz,Kriechbaum_PRB84} Since we focus on heterostructures grown in [111] direction, we choose the $z$ axis parallel to [111]. The Hamiltonian in the basis of $\frac{1}{\sqrt{2}}(X-iY)\ket{\uparrow}$, $\frac{1}{\sqrt{2}}(X+iY)\ket{\downarrow}$,$\frac{1}{\sqrt{2}}(S_X-iS_Y)\ket{\uparrow}$, $\frac{1}{\sqrt{2}}(S_X+iS_Y)\ket{\downarrow}$ (where $S_{X,Y}$ transform like $X,Y$ but do not change sign under reflection) is then given by
\begin{equation}
\hat{H} = \epsilon(\mathbf{k}) + \left(
   \begin{array}{cc}
     \mathcal{M}(\mathbf{k}) &  \hat{\bf\sigma} \cdot {\bf Q} \\
      \hat{\bf\sigma}\cdot {\bf Q}  & -\mathcal{M}(\mathbf{k}) \\
   \end{array}
 \right) + \hat{H}_{S},\qquad  \label{eq:H}
\end{equation}
where $\epsilon(\mathbf{k}) \! = \! V(x)+D_{1}k^{2}_{z}+D_{2}k^{2}_{\bot}$, $\mathcal{M}(\mathbf{k})\! = \! \Delta(x) +  B_{1}k^{2}_{z}+B_{2}k^{2}_{\bot}$ and ${\bf Q}\! = \! \mathbf{\hat{P}_{0}}\cdot \mathbf{k}$. For [111] valley $\mathbf{\hat{P}_{0}}$ is a matrix with $(v_\bot,v_\bot,v_\|)$ on the diagonal. We stress that a realistic large ratio of $v_{\bot}/v_{\|}$ is crucial for our considerations. 
For remaining oblique valleys we replace $\mathbf{\hat{P}_{0}}$ with $\mathbf{\hat{P}}\!  = \! \mathbf{\hat{S}} \mathbf{\hat{P}_{0}}\mathbf{\hat{S}}^{-1}$, where $\mathbf{\hat{S}}$ is a transformation matrix between the coordinate systems associated with the oblique valley $\{\hat{\mathbf{e}}_{\alpha'}\}$, with $\hat{\mathbf{e}}_{z'}$ along the valley direction, and $\{\hat{\mathbf{e}}_{\alpha}\}$, with $\hat{\mathbf{e}}_{z}$ along the growth axis: $(\hat{\mathbf{e}}_{x'},\hat{\mathbf{e}}_{y'},\hat{\mathbf{e}}_{z'}) = (\hat{\mathbf{e}}_{x},\hat{\mathbf{e}}_{y},\hat{\mathbf{e}}_{z})\cdot \mathbf{\hat{S}}$.

We take $\Delta(x) \! = \! 0.095-0.26x$ eV, corresponding to the gap inversion at $x_{c}\! \approx \! 0.37$, and $V(x) \! = \! 0.125 x$ eV.\cite{Kriechbaum_PRB84} For the remaining parameters we take $D_{1(2)} \! = \! 1.65 \, (9.25)$ eV\AA$^2$, $B_{1(2)} \! = \! 6.15\, (49.35)$ eV\AA$^2$, $v_{\|} \! = \! 1.44$ eV\AA, and $v_{\bot} \! = \! 4.76$ eV\AA{}, and  we neglect their composition which is, if any, very weak in the considered regime of $x\! < 0.5$. Fitting of the Hamiltonian parameters to the results of Nuclear Magnetic Resonance\cite{Hewes_PRB73} measurements in \PbSnTe with $x$ up to $0.6$ showed no changes in $v_{\bot}$ and $v_\|$ parameters, while the $x$ dependence of $B_{1(2)}$ and $D_{1(2)}$ was inferred to be very weak. In other works\cite{Appold79} a constant $v_{\bot}$ was inferred for $x$ up to $0.16$, while $v_\|$ was seen to change by an amount which extrapolates to about $16$ \% change for $x\! = \! 0.46$ compared to PbTe. Experimental data on the $v_{\bot}/v_{\|}$ ratio collected in Ref.~\onlinecite{He_JPF85} also suggests a very weak dependence of these parameters on Sn content up to $x\! \approx \! 0.5$.

The strain described by tensor $\hat{\epsilon}$ is accounted for by $\hat{H}_{S} = \text{diag}\{ D^{c}_{d}\text{Tr}\hat{\epsilon} + D^{c}_{u}\epsilon_{zz},D^{v}_{d}\text{Tr}\hat{\epsilon} + D^{v}_{u}\epsilon_{zz}\}$, in which the two entries correspond to $2\times 2$ blocks. 
Note that $\hat{H}_{S}$ gives strain-dependent corrections to $\Delta$ and $V$.
The acoustic deformation potentials are taken again as composition-independent: $D^{c}_{d} \! = \! -1.09$ eV, $D^{v}_{d} \! = \! -2.23$ eV, $D^{c}_{u} \! = \! 2.07$ eV, and $D^{v}_{u} \! = \! 2.62$ eV.\cite{Kriechbaum_PRB84} The lattice constants of PbTe (SnTe) are taken as $a_{0} \! = \!6.454$ \AA {} ($6.313$ \AA).

The above Hamiltonian is written in a form which allows one to immediately see that after a basis reordering it becomes the same as the recently studied model $\mathbf{k}\!\cdot\!\mathbf{p}$ Hamiltonian describing the states near the $\Gamma$ point in the Bi$_{2}$Se$_{3}$ family of materials.\cite{Liu_model_PRB10} There, the TI phase occurs\cite{Liu_model_PRB10,Liu_crossover_PRB10} when $\Delta\cdot B_{1,2} \! < \! 0$, i.e.~when the conduction  and valence bands at $\Gamma$ point become inverted. As mentioned before, in unstrained \PbSnTe the inversion occurs simultaneously at four L points, and the bulk phase remains topologically trivial.

\section{Helical nature of interface states.}  \label{sec:helical}
We first analyze interface-bound states (IBS) localized at two decoupled heterointerfaces of a very wide PbTe/\PbSnTe heterostructure grown in [111] direction. We assume that for $|z|\! < \! d/2$ we have the values of $\Delta \! > \! 0$ and $V$ corresponding to PbTe, while for $|z| \! > \! d/2$, $\Delta'\! < \! 0$ and $V'$ correspond to Pb$_{0.54}$Sn$_{0.46}$Te. From the point of view of band lineup, this structure is in fact a finite width potential step - a quantum \emph{wall} (QWa) instead of a quantum well (QWe). 

We use the effective mass approximation and replace $k_{z}$ by $-i\partial/\partial z$ in Eq.~(\ref{eq:H}). As long as the intervalley coupling (due to alloy disorder or atomic-scale interface reconstruction) can be neglected, the presence of mutual band inversion at the interfaces guarantees the existence of helical Dirac-like IBS.\cite{Volkov_JETP85,Pankratov_SST90} The dispersion and the spin structure of IBS is insensitive to exact profile of the interface,\cite{Volkov_JETP85,Pankratov_SST90} with only the $z$ dependence of wavefunctions being affected in the region of interdiffusion. For simplicity we assume an abrupt interface,\cite{Korenman_PRB87,Agassi_PRB88,Dugaev_pssb94} and use the boundary conditions of continuity of the wavefunction and its derivative (the latter only when we consider nonzero $k^{2}$ terms). Note that these boundary conditions are sufficient in the case when, as we assume, the $\mathbf{k}\cdot\mathbf{p}$ Hamiltonian's parameters such as $v_\bot$, $v_\|$, $D_{1,2}$ and $B_{1,2}$ are the same for both compounds.

Since our numerical calculations show that for realistic parameters the $k^{2}$ terms are of minor quantitative importance, we present much simpler analytical results  obtained by neglecting them. We also focus on results for [111] valley, which we label with superscript $\alpha$ (with $\beta$ denoting the oblique-valley results).
The energies of the IBS are $E^{\alpha}_{\pm}(k_\bot ) \! = \! E_{0} \pm v_\bot k_\bot \frac{\mathcal{D}}{\Delta-\Delta'}$, where $\mathcal{D}^2 \! = \! (\Delta-\Delta')^2-(V-V')^2$ and $E_{0} \! = \! \frac{\Delta V'-V\Delta'}{\Delta-\Delta'}$ (in oblique valleys the dispersion is of course anisotropic in $\mathbf{k}_{\perp}$).
The eigenvectors are 4-spinors $\hat{\psi}_{\pm}^{L/R}(\mathbf{k}_{\bot})$, with $L/R$ denoting the left ($z\! = \! -d/2$) and the right ($z\! = \! d/2$) interfaces, multiplied by appropriate functions $f^{L/R}_{\pm}(z)$ describing the exponential decay away from a given interface. The inverse decay lengths are given by 
\beq
\chi_{\pm}^{\alpha} \! = \! \frac{1}{v_\|}(\mathcal{D}\frac{\Delta}{\Delta-\Delta'} \pm \eta v_\bot k_\bot) \,\, \label{eq:chi}
\eeq  in the middle layer and $\chi'_{\mp} \! = \! \chi_{\pm} (\Delta \leftrightarrow \Delta')$ outside, with $\eta \! \equiv \! (V-V')/(\Delta-\Delta')$. Let us note that in the oblique valley case $\chi^{\beta}_{\pm}$ acquire an imaginary part at finite $\mathbf{k}_{\bot}$ - however, what is crucial is the fact that at $k_{\bot}\!=\!0$, $\chi^\beta_{0}$ is given by Eq.~(\ref{eq:chi}) with $v_{\|}$ replaced by $\frac{1}{3}\sqrt{8v^{2}_{\bot}+v^{2}_{\|}}$.
 
The two-dimensional subspaces of IBS at the two interfaces are spanned by 4-spinors   $\{\hat{\phi}^{L/R}_{\uparrow},\hat{\phi}^{L/R}_{\downarrow}\}$, for which the subscripts signify that their components correspond to well-defined spin states, although they remain linear combinations of different orbitals. Defining $\gamma \! \equiv \! \arctan \sqrt{(1+\eta)/(1-\eta)}$ we can write them as: $\hat{\phi}^{L/R}_{\uparrow} \! = \! [ \mp i\cos\gamma,0,\sin\gamma,0]^{T}$, $\hat{\phi}^{L/R}_{\downarrow} \! = \! [0,\pm i\cos\gamma,0,\sin\gamma ]^{T}$. Using these basis states we can write the 4-spinor parts of the IBS solutions: $\hat{\psi}_{\pm}^{L}(\mathbf{k}_{\bot}) \! = \! (\pm i \hat{\phi}^{L}_{\uparrow} + e^{i\theta} \hat{\phi}^{L}_{\downarrow})/\sqrt{2}$, and $\hat{\psi}_{\pm}^{R}(\mathbf{k}_{\bot}) \! = \! ( \hat{\phi}^{R}_{\uparrow} \pm i e^{i\theta} \hat{\phi}^{R}_{\downarrow})/\sqrt{2}$, where $e^{i\theta} \! =\! (k_{x}+ik_{y})/k_{\bot}$.

The helical nature of these states can be seen by calculating the expectation values of spin operators: $\bra{ \hat{\psi}_{\pm}^{L}}\hat{\sigma}_{x}\ket{\hat{\psi}_{\pm}^{L}} \! = \! \mp \cos 2\gamma \sin \theta$, $\bra{ \hat{\psi}_{\pm}^{L}}\hat{\sigma}_{y}\ket{\hat{\psi}_{\pm}^{L}} \! = \! \pm \cos 2\gamma \cos \theta$, i.e.~the spin vector is perpendicular to $\mathbf{k}_{\bot}$, pointing clockwise (anticlockwise) for negative (positive) energy branch. The pattern on the $R$ interface is reversed. 
More generally \cite{cosgamma_comment} we can write an effective surface Hamiltonian in the $\{\hat{\phi}^{L/R}_{\uparrow},\hat{\phi}^{L/R}_{\downarrow}\}$ basis:
\beq
\hat{H}^{L/R}_{\text{surf}} = \pm v_\bot \frac{\mathcal{D}}{\Delta-\Delta'} ( \tilde{\sigma}_{x}k_{y} - \tilde{\sigma}_{y}k_{x} ) \,\, , \label{eq:H2}
\eeq
with the $\tilde{\sigma}$ Pauli matrices operating in the respective two-dimensional spaces. 
Let us remark that the spin pattern for states from the oblique valleys is more complicated, with nonzero out-of-plane $\langle \hat{\sigma}_{z}\rangle$  polarization.

All these results show complete analogy with the surface states in TIs such as Bi$_{2}$Se$_{3}$.\cite{Liu_model_PRB10} The difference is the existence of four pairs of IBS, with the states from all the valleys coexisting at a given energy in the bulk gap, showing that the PbTe/\PbSnTe heterointerface is analogous to a surface of a weak TI. 

\begin{figure}[t]
\includegraphics[width = 0.9\linewidth]{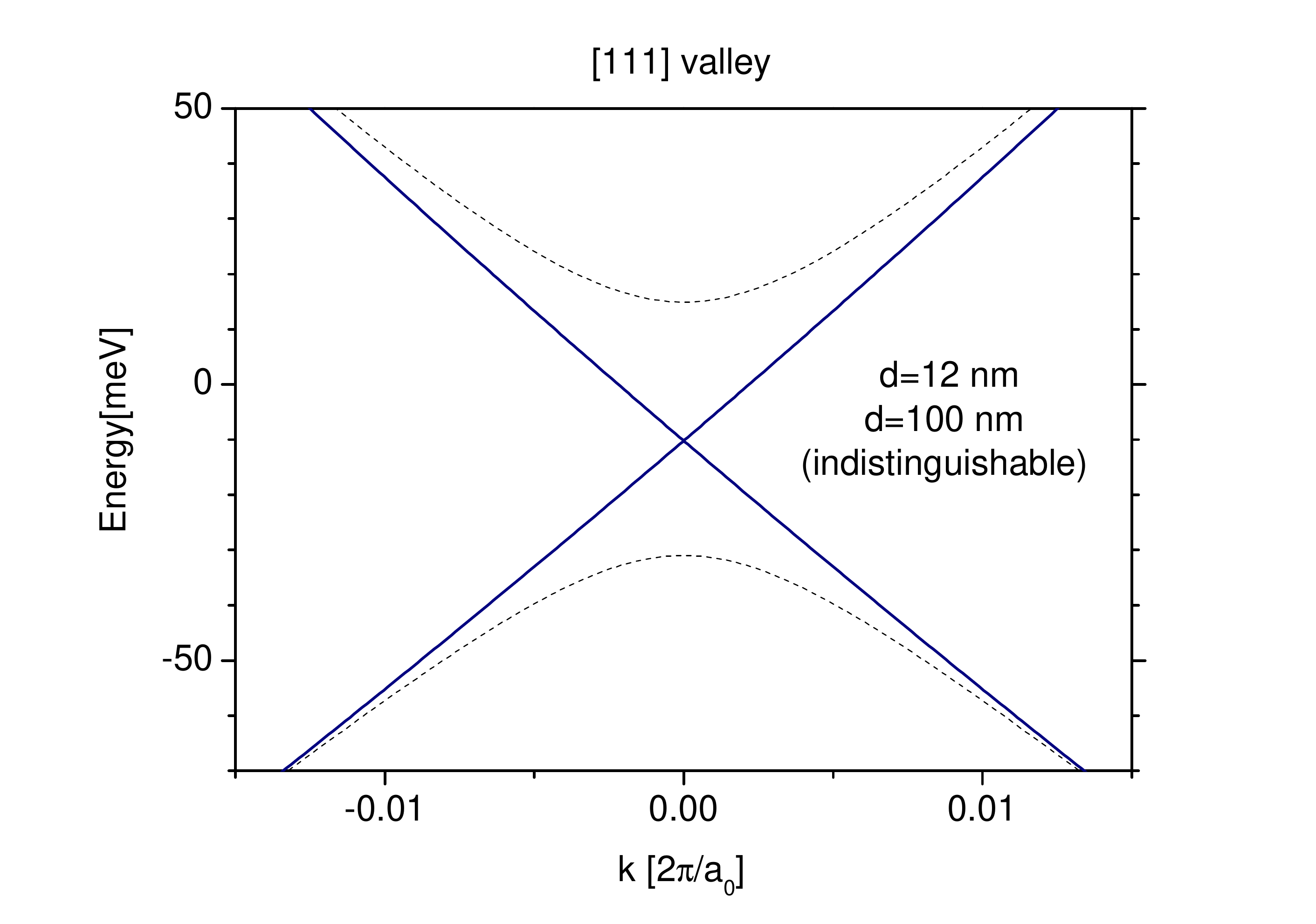}
\caption{(Color online) Solid line is the dispersion of the [111] valley interface bound states for PbTe/Pb$_{0.54}$Sn$_{0.46}$Te quantum wall grown in [111] direction, with PbTe widths $d$ of 12 and 100 nm (undistinguishable), plotted as a function of in-plane $k$ vector.
The dashed lines are the bulk band edges for Pb$_{0.54}$Sn$_{0.46}$Te (the band edges of PbTe are outside of the shown energy region). There is a gap of $0.2$ meV opened at the Dirac point for $d\! = \! 12$ nm, see Fig.~\ref{fig:Eg}.}
\label{fig:E111}
\end{figure}

\begin{figure}[t]
\includegraphics[width = 0.9\linewidth]{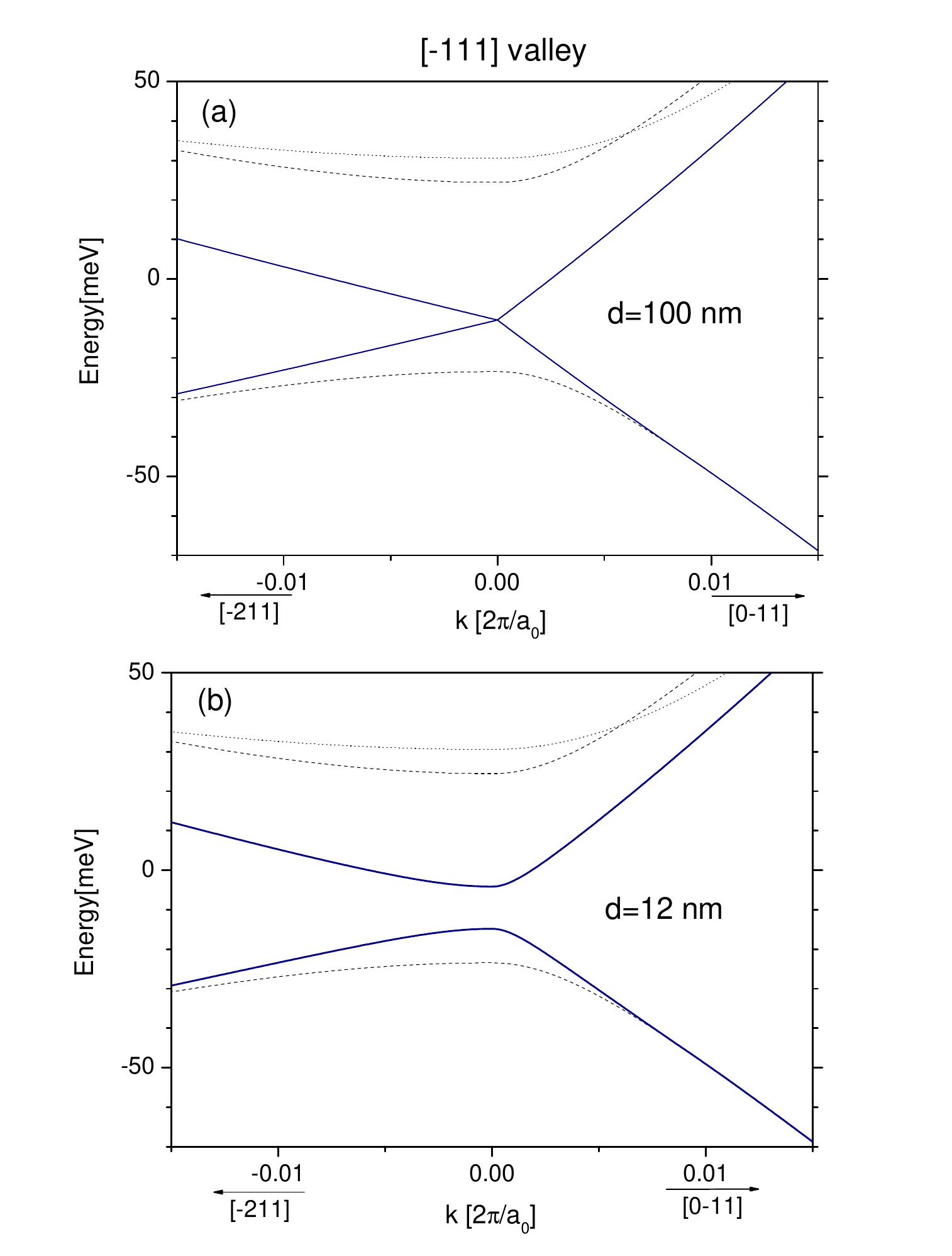}
\caption{(Color online) Solid line is the dispersion of the $[\bar{1}11]$ valley interface-bound states for PbTe/Pb$_{0.54}$Sn$_{0.46}$Te quantum wall with PbTe width $d$ of (a) 100 nm  and  (b) 12 nm. The dashed (dotted) lines are the bulk band edges for Pb$_{0.54}$Sn$_{0.46}$Te (PbTe). The gap of $\sim\! 10$ meV opens for $d \! =\! 12$ nm.}
\label{fig:E_oblique}
\end{figure}

\section{The effect of finite Quantum Wall width.}  \label{sec:width}
We assume that the lattice mismatch between PbTe and Pb$_{0.54}$Sn$_{0.46}$Te is shared in $4\! :\! 1$ ratio between the two materials (i.e.~assuming a common lattice constant corresponding to $0.37$ content of Sn).
This requires an engineering of strain distribution by, e.g., choice of a substrate. 
If we neglect the $k^2$ terms the calculation is fairly straightforward.\cite{Korenman_PRB87,Agassi_PRB88,Dugaev_pssb94} With the $k^2$ terms and for a general valley direction we have to resort to solving the problem numerically.

\begin{figure}[t]
\includegraphics[width = \linewidth]{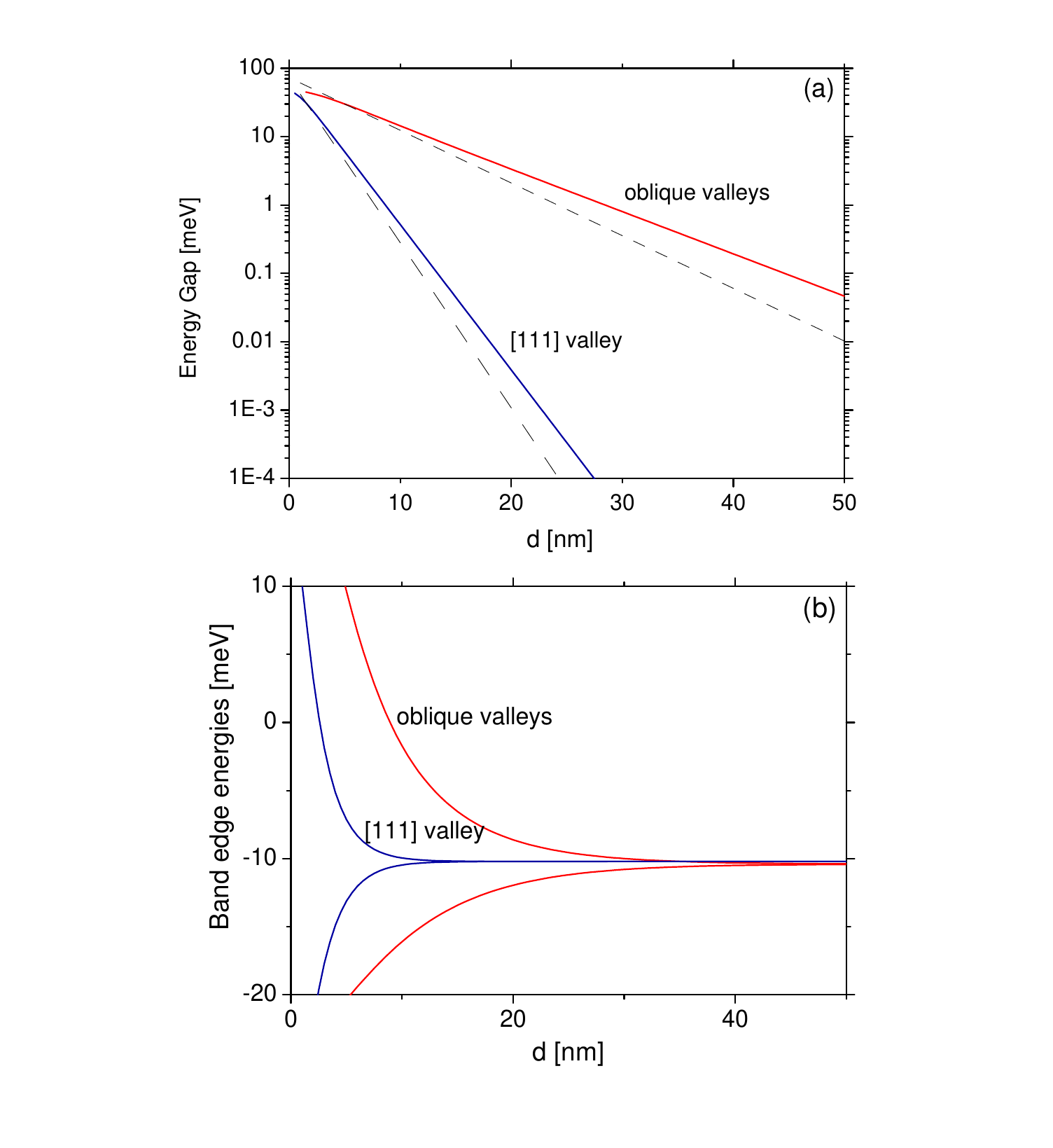}
\caption{ (Color online) (a) The energy gaps of the IBS-derived states from [111] valley and the three oblique valleys. The solid lines are the results of an exact calculation, while the dashed lines are obtained with Eq.~(\ref{eq:Eg}), which neglects the $k^{2}$ terms in the Hamiltonian. (b) Band edge energies for IBS-derived states. The Dirac point of [111] valley states is located in the gap of the states from the remaining valleys. }
\label{fig:Eg}
\end{figure}

In Fig.~\ref{fig:E111} we show the energy dispersion of the [111] valley states formed in the QWa. For both widths of $d\!=\! 12$ and $100$ nm the IBS from the two interfaces remain practically uncoupled: the gap opened for $d\!=\! 12$ nm is only $\approx\! 0.2$ meV. On the other hand, as we show in Fig.~\ref{fig:E_oblique}, with decreasing $d$ a sizable gap opens for the interface bound (IB) derived states (formed due to the coupling of IBS from the opposite interfaces) in the oblique valleys, and it is $\sim\! 10$ meV for $d\! = \! 12$ nm. 
As shown in Fig.~\ref{fig:Eg}a, the gaps $E_{g}^{\alpha/\beta}$ are visibly different. Thus, for most of practical purposes (e.g.~for experiments done at temperature $\sim \! 1$ K), in a structure with $d~\! \approx \! 10$ nm the dispersion of the states from the [111] valley can be considered gapless, while the states from the oblique valleys are gapped. Let us also stress that at energies $|E_{\pm}(\mathbf{k}_{\perp})-E_{0}| \! \gg \! E_{g}$, the eigenstates are essentially the L/R interface bound states (the description in terms of massive Dirac fermions \cite{Lu_PRB10} is relevant only at small $\mathbf{k}_{\perp}$).
Furthermore, as shown in Fig.~\ref{fig:Eg}b, for the chosen QWa the [111] valley states are in the gap of the oblique valley solutions, so that disorder with typical energy scale smaller than this gap cannot mix the states from [111] valley with the others. Thus, for epitaxial interface of good quality we expect that the above-described system is basically equivalent to a  layer of material such as Bi$_{2}$Te$_{3}$ or Bi$_{2}$Se$_{3}$, with the only difference being the smaller size of the gap within which the surface states exist ($\sim 10$ meV vs $\sim 0.3$ eV).

The above results can be understood using a calculation with the $k^{2}$ terms neglected and $d$ assumed to be large. We orthogonalize the IBS and form their symmetric and antisymmetric combinations, approximating the exact solutions in large $d$ limit. We calculate the gap by evaluating the matrix elements of the QWa Hamiltonian between them. The $d$-dependence of $E_{g}^{\alpha,\beta}$ comes from exponential decay of the IBS in the step region, described by  inverse lengths $\chi^{\alpha,\beta}_{0}$ (calculated for $k_\bot \! =\! 0$) given in Eq.~(\ref{eq:chi}) and below.
In this way we obtain the bandgaps (see also Ref.~\onlinecite{Korenman_PRB87})
\beq
E^{\alpha/\beta}_{g} = 4\mathcal{D}^{2}\frac{|\Delta\Delta'|}{(\Delta-\Delta')^{3}}\exp(-\chi^{\alpha/\beta}d) \,\, . \label{eq:Eg}
\eeq 
The predictions of this formula agree well with the results of the full calculation shown in Fig.~\ref{fig:Eg}a. 

\section{Discussion.} \label{sec:discussion}
\begin{figure}[t]
\includegraphics[width = 0.9\linewidth]{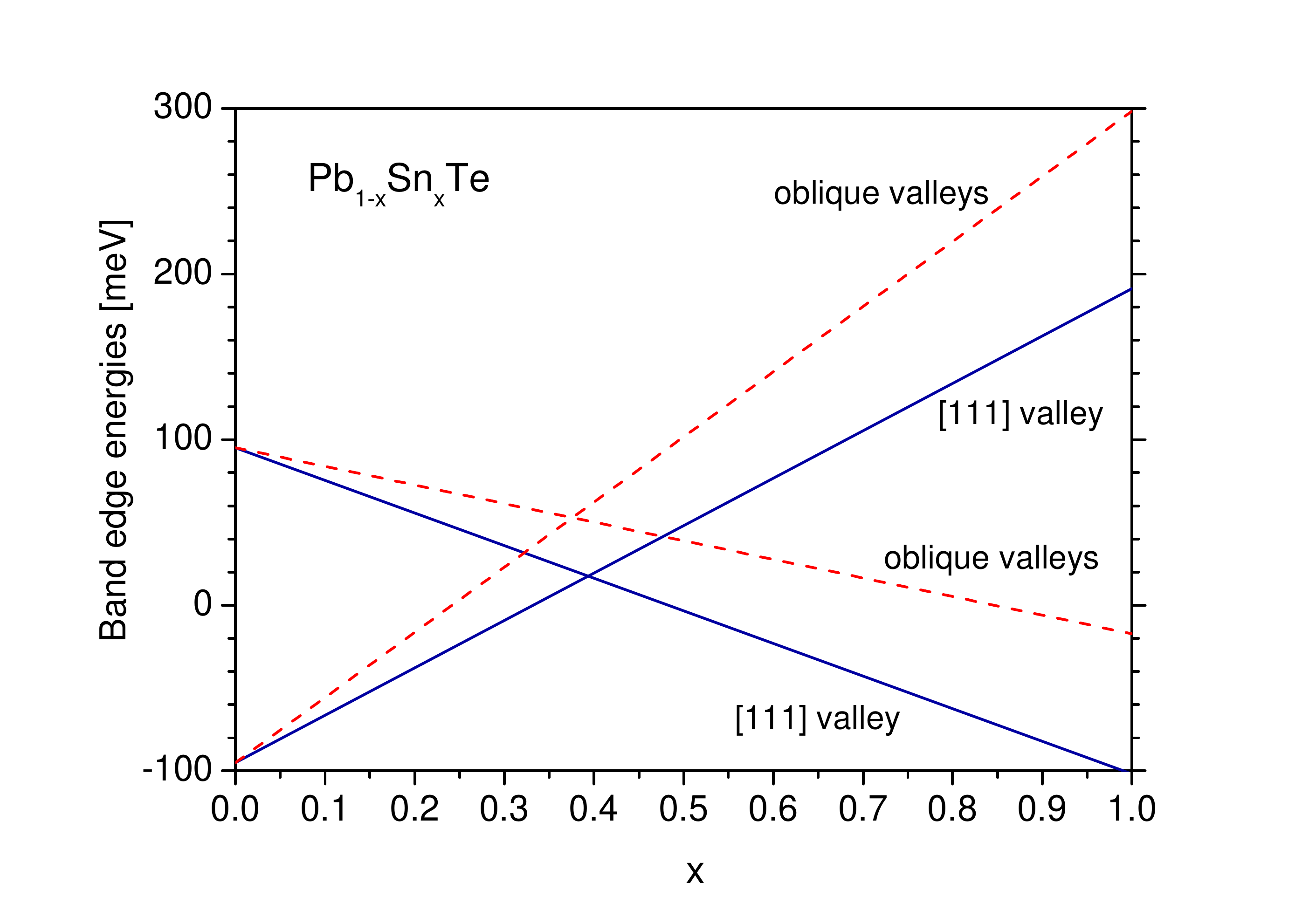}
\caption{(Color online) The energies of extrema of conduction and valence bands for \PbSnTe lattice matched to PbTe (grown in [111] direction), calculated using Eq.~(\ref{eq:H}) as a function of $x$. }
\label{fig:strain}
\end{figure}

The above choice of a heterostructure was dictated by strain dependence of energies of band extrema in Pb$_{1-x}$Sn$_{x}$Te. As shown in Fig.~\ref{fig:strain}, when the lattice constant of \PbSnTe is adjusted to the PbTe constant in a heterostructure grown in [111] direction, with the realistic values of deformation potentials \cite{Kriechbaum_PRB84} we obtain that the gaps at [111] L point and the remaining ones. 
One can see then that in the \PbSnTe quantum well (with PbTe being the barrier), it is practically impossible (without applying a compressive strain) to obtain a Dirac point of [111] valley IBS within the gap of the oblique valley IB-derived states. For the QWa with material parameters adopted here,  we only require a modest amount of additional tensile strain, which could be provided at low temperatures by BaF$_{2}$ substrate typically used for IV-VI compounds.\cite{Kriechbaum_PRB84} Additional tuning of band edge offsets can also be obtained by using quanternary alloys and replacing a fraction of Te by Se.

Note however that if band-inverted \PbSnTe itself possesses gapless surface states,\cite{Hsieh_arXiv12} which would then appear at its interfaces with the substrate and with the gate covering the structure, then the choice of a compressively strained \PbSnTe/PbTe quantum well will be more experimentally practical. In such a case, apart from the IB-derived states we obtain also the ``normal'' quantum well states, with energies outside of the gap of \PbSnTe. The behavior of gaps of IB-derived states as a function of $d$ is analogous to the QWa case, and, provided that the appropriate compressive strain is aplied to the structure, the Dirac point of the [111] valley states can be located in the gap of the remaining IB-derived states (and, of course, also also in the gap of all the ``normal'' quantum well states).

Let us note that a result qualitatively the same as shown in Fig.~\ref{fig:strain} can be obtained for \PbSnTe uniaxially strained in [111] direction. The non-overlapping of bandgaps in the range of $x$ corresponding the inverted oblique valley and non-inverted [111] valley implies that applying such a strain, as proposed in Ref.~\onlinecite{Fu_PRB07}, cannot change \PbSnTe into a strong TI, but into a topological semimetal.

Another possible practical obstacle in investigation of helical states in the proposed structure is the issue of p-type self-doping\cite{Prokofieva_Semiconductors10} of \PbSnTe with $x \! > \! 0.2$. 
This is analogous to the situation in materials from Bi$_{2}$Se$_{3}$ family, the members of which possess large bulk conductivity when as-grown, see e.g.~Ref.~\onlinecite{Butch_PRB10}. At $x\! = \! 0.25$ holes were compensated by indium doping,\cite{Khokhlov_APL00} but it remains to be investigated how effective this would be at $x \! \sim \! 0.4$ considered here.

\section{Conclusions.} We have calculated the bandstructure of a heterostructure  based on PbTe and band-inverted Pb$_{1-x}$Sn$_{x}$Te{}, in which PbTe forms a  quantum wall (QWa) of finite thickness $d$. The ``normal'' quantum well states are then absent, and the localized solutions appear due to the mutual band inversion between the constituent materials. For very large $d$ these solutions correspond to four pairs of Dirac cones located at the two interfaces.\cite{Volkov_JETP85,Korenman_PRB87,Agassi_PRB88,Pankratov_SST90} As the QWa grown in the [111] direction is narrowed, these states become gapped due to the overlap of wavefunctions localized on the two interfaces. 
We have found that the opened gap is at least two orders of magnitude larger for states associated with three oblique valleys compared to the states from the [111] valley. Thus, after properly choosing the QWa composition, its width, and the strain distribution in it, one can obtain a practically gapless pair of helical Dirac states from the [111] valley, with their Dirac point located in the gap of the states from the other valleys. The IV-VI heterostructure can then serve as an analogue of a thin layer of strong topological insulator.

\section{Acknowledgements.} We thank T.~Dietl, J.~Tworzyd{\l}o, G.~Springholz, T.~Story and G.~Grabecki for discussions.
We acknowledge support from Iuventus Plus grant (0060/H03/2010/70) of the Polish Ministry of Science, from EC network SemiSpinNet (PITN-GA-2008-215368), and from the EU FunDMS Advanced
Grant of the European Research Council within the "Ideas" Seventh Framework Programme.


\begin{thebibliography}{0}
\expandafter\ifx\csname natexlab\endcsname\relax\def\natexlab#1{#1}\fi
\expandafter\ifx\csname bibnamefont\endcsname\relax
  \def\bibnamefont#1{#1}\fi
\expandafter\ifx\csname bibfnamefont\endcsname\relax
  \def\bibfnamefont#1{#1}\fi
\expandafter\ifx\csname citenamefont\endcsname\relax
  \def\citenamefont#1{#1}\fi
\expandafter\ifx\csname url\endcsname\relax
  \def\url#1{\texttt{#1}}\fi
\expandafter\ifx\csname urlprefix\endcsname\relax\def\urlprefix{URL }\fi
\providecommand{\bibinfo}[2]{#2}
\providecommand{\eprint}[2][]{\url{#2}}

\end{thebibliography}


\begin{thebibliography}{45}
\expandafter\ifx\csname natexlab\endcsname\relax\def\natexlab#1{#1}\fi
\expandafter\ifx\csname bibnamefont\endcsname\relax
  \def\bibnamefont#1{#1}\fi
\expandafter\ifx\csname bibfnamefont\endcsname\relax
  \def\bibfnamefont#1{#1}\fi
\expandafter\ifx\csname citenamefont\endcsname\relax
  \def\citenamefont#1{#1}\fi
\expandafter\ifx\csname url\endcsname\relax
  \def\url#1{\texttt{#1}}\fi
\expandafter\ifx\csname urlprefix\endcsname\relax\def\urlprefix{URL }\fi
\providecommand{\bibinfo}[2]{#2}
\providecommand{\eprint}[2][]{\url{#2}}

\bibitem[{\citenamefont{Hasan and Kane}(2010)}]{Hasan_RMP10}
\bibinfo{author}{\bibfnamefont{M.~Z.} \bibnamefont{Hasan}} \bibnamefont{and}
  \bibinfo{author}{\bibfnamefont{C.~L.} \bibnamefont{Kane}},
  \bibinfo{journal}{Rev. Mod. Phys.} \textbf{\bibinfo{volume}{82}},
  \bibinfo{pages}{3045} (\bibinfo{year}{2010}).

\bibitem[{\citenamefont{Qi and Zhang}(2011)}]{Qi_RMP11}
\bibinfo{author}{\bibfnamefont{X.-L.} \bibnamefont{Qi}} \bibnamefont{and}
  \bibinfo{author}{\bibfnamefont{S.-C.} \bibnamefont{Zhang}},
  \bibinfo{journal}{Rev. Mod. Phys.} \textbf{\bibinfo{volume}{83}},
  \bibinfo{pages}{1057} (\bibinfo{year}{2011}).

\bibitem[{\citenamefont{Konig et~al.}(2007)\citenamefont{Konig, Wiedmann,
  Brune, Roth, Buhmann, Molenkamp, Qi, and Zhang}}]{Konig_Science07}
\bibinfo{author}{\bibfnamefont{M.}~\bibnamefont{Konig}},
  \bibinfo{author}{\bibfnamefont{S.}~\bibnamefont{Wiedmann}},
  \bibinfo{author}{\bibfnamefont{C.}~\bibnamefont{Brune}},
  \bibinfo{author}{\bibfnamefont{A.}~\bibnamefont{Roth}},
  \bibinfo{author}{\bibfnamefont{H.}~\bibnamefont{Buhmann}},
  \bibinfo{author}{\bibfnamefont{L.~W.} \bibnamefont{Molenkamp}},
  \bibinfo{author}{\bibfnamefont{X.-L.} \bibnamefont{Qi}}, \bibnamefont{and}
  \bibinfo{author}{\bibfnamefont{S.-C.} \bibnamefont{Zhang}},
  \bibinfo{journal}{Science} \textbf{\bibinfo{volume}{318}},
  \bibinfo{pages}{766} (\bibinfo{year}{2007}).

\bibitem[{\citenamefont{Knez et~al.}(2011)\citenamefont{Knez, Du, and
  Sullivan}}]{Knez_PRL11}
\bibinfo{author}{\bibfnamefont{I.}~\bibnamefont{Knez}},
  \bibinfo{author}{\bibfnamefont{R.-R.} \bibnamefont{Du}}, \bibnamefont{and}
  \bibinfo{author}{\bibfnamefont{G.}~\bibnamefont{Sullivan}},
  \bibinfo{journal}{Phys.\ Rev.\ Lett.} \textbf{\bibinfo{volume}{107}},
  \bibinfo{pages}{136603} (\bibinfo{year}{2011}).

\bibitem[{\citenamefont{Fu and Kane}(2007)}]{Fu_PRB07}
\bibinfo{author}{\bibfnamefont{L.}~\bibnamefont{Fu}} \bibnamefont{and}
  \bibinfo{author}{\bibfnamefont{C.~L.} \bibnamefont{Kane}},
  \bibinfo{journal}{Phys.\ Rev.\ B} \textbf{\bibinfo{volume}{76}},
  \bibinfo{pages}{045302} (\bibinfo{year}{2007}).

\bibitem[{\citenamefont{Fu and Kane}(2008)}]{Fu_PRL08}
\bibinfo{author}{\bibfnamefont{L.}~\bibnamefont{Fu}} \bibnamefont{and}
  \bibinfo{author}{\bibfnamefont{C.~L.} \bibnamefont{Kane}},
  \bibinfo{journal}{Phys.\ Rev.\ Lett.} \textbf{\bibinfo{volume}{100}},
  \bibinfo{pages}{096407} (\bibinfo{year}{2008}).

\bibitem[{\citenamefont{Qi et~al.}(2008)\citenamefont{Qi, Hughes, and
  Zhang}}]{Qi_PRB08}
\bibinfo{author}{\bibfnamefont{X.-L.} \bibnamefont{Qi}},
  \bibinfo{author}{\bibfnamefont{T.~L.} \bibnamefont{Hughes}},
  \bibnamefont{and} \bibinfo{author}{\bibfnamefont{S.-C.} \bibnamefont{Zhang}},
  \bibinfo{journal}{Phys.\ Rev.\ B} \textbf{\bibinfo{volume}{78}},
  \bibinfo{pages}{195424} (\bibinfo{year}{2008}).

\bibitem[{\citenamefont{Essin et~al.}(2009)\citenamefont{Essin, Moore, and
  Vanderbilt}}]{Essin_PRL09}
\bibinfo{author}{\bibfnamefont{A.~M.} \bibnamefont{Essin}},
  \bibinfo{author}{\bibfnamefont{J.~E.} \bibnamefont{Moore}}, \bibnamefont{and}
  \bibinfo{author}{\bibfnamefont{D.}~\bibnamefont{Vanderbilt}},
  \bibinfo{journal}{Phys.\ Rev.\ Lett.} \textbf{\bibinfo{volume}{102}},
  \bibinfo{pages}{146805} (\bibinfo{year}{2009}).

\bibitem[{\citenamefont{Xia et~al.}(2009)\citenamefont{Xia, Qian, Hsieh, Wray,
  Pal, Lin, Bansil, Grauer, Hor, Cava et~al.}}]{Xia_NP09}
\bibinfo{author}{\bibfnamefont{Y.}~\bibnamefont{Xia}},
  \bibinfo{author}{\bibfnamefont{D.}~\bibnamefont{Qian}},
  \bibinfo{author}{\bibfnamefont{D.}~\bibnamefont{Hsieh}},
  \bibinfo{author}{\bibfnamefont{L.}~\bibnamefont{Wray}},
  \bibinfo{author}{\bibfnamefont{A.}~\bibnamefont{Pal}},
  \bibinfo{author}{\bibfnamefont{H.}~\bibnamefont{Lin}},
  \bibinfo{author}{\bibfnamefont{A.}~\bibnamefont{Bansil}},
  \bibinfo{author}{\bibfnamefont{D.}~\bibnamefont{Grauer}},
  \bibinfo{author}{\bibfnamefont{Y.~S.} \bibnamefont{Hor}},
  \bibinfo{author}{\bibfnamefont{R.~J.} \bibnamefont{Cava}},
  \bibnamefont{et~al.}, \bibinfo{journal}{Nat. Phys.}
  \textbf{\bibinfo{volume}{5}}, \bibinfo{pages}{398} (\bibinfo{year}{2009}).

\bibitem[{\citenamefont{Chen et~al.}(2009)\citenamefont{Chen, Analytis, Chu,
  Liu, Mo, Qi, Zhang, Lu, Dai, Fang et~al.}}]{Chen_Science09}
\bibinfo{author}{\bibfnamefont{Y.~L.} \bibnamefont{Chen}},
  \bibinfo{author}{\bibfnamefont{J.~G.} \bibnamefont{Analytis}},
  \bibinfo{author}{\bibfnamefont{J.-H.} \bibnamefont{Chu}},
  \bibinfo{author}{\bibfnamefont{Z.~K.} \bibnamefont{Liu}},
  \bibinfo{author}{\bibfnamefont{S.-K.} \bibnamefont{Mo}},
  \bibinfo{author}{\bibfnamefont{X.~L.} \bibnamefont{Qi}},
  \bibinfo{author}{\bibfnamefont{H.~J.} \bibnamefont{Zhang}},
  \bibinfo{author}{\bibfnamefont{D.~H.} \bibnamefont{Lu}},
  \bibinfo{author}{\bibfnamefont{X.}~\bibnamefont{Dai}},
  \bibinfo{author}{\bibfnamefont{Z.}~\bibnamefont{Fang}}, \bibnamefont{et~al.},
  \bibinfo{journal}{Science} \textbf{\bibinfo{volume}{325}},
  \bibinfo{pages}{178} (\bibinfo{year}{2009}).

\bibitem[{\citenamefont{Checkelsky et~al.}(2011)\citenamefont{Checkelsky, Hor,
  Cava, and Ong}}]{Checkelsky_PRL11}
\bibinfo{author}{\bibfnamefont{J.~G.} \bibnamefont{Checkelsky}},
  \bibinfo{author}{\bibfnamefont{Y.~S.} \bibnamefont{Hor}},
  \bibinfo{author}{\bibfnamefont{R.~J.} \bibnamefont{Cava}}, \bibnamefont{and}
  \bibinfo{author}{\bibfnamefont{N.~P.} \bibnamefont{Ong}},
  \bibinfo{journal}{Phys. Rev. Lett.} \textbf{\bibinfo{volume}{106}},
  \bibinfo{pages}{196801} (\bibinfo{year}{2011}).

\bibitem[{\citenamefont{Sacepe et~al.}(2011)\citenamefont{Sacepe, Oostinga, Li,
  Ubaldini, Couto, Giannini, and Morpurgo}}]{Sacepe_NC11}
\bibinfo{author}{\bibfnamefont{B.}~\bibnamefont{Sacepe}},
  \bibinfo{author}{\bibfnamefont{J.~B.} \bibnamefont{Oostinga}},
  \bibinfo{author}{\bibfnamefont{J.}~\bibnamefont{Li}},
  \bibinfo{author}{\bibfnamefont{A.}~\bibnamefont{Ubaldini}},
  \bibinfo{author}{\bibfnamefont{N.~J.~G.} \bibnamefont{Couto}},
  \bibinfo{author}{\bibfnamefont{E.}~\bibnamefont{Giannini}}, \bibnamefont{and}
  \bibinfo{author}{\bibfnamefont{A.~F.} \bibnamefont{Morpurgo}},
  \bibinfo{journal}{Nat.~Comm.} \textbf{\bibinfo{volume}{2}},
  \bibinfo{pages}{575} (\bibinfo{year}{2011}).

\bibitem[{\citenamefont{Kim et~al.}(2012)\citenamefont{Kim, Cho, Butch, Syers,
  Kirshenbaum, Adam, Paglione, and Fuhrer}}]{Kim_NP12}
\bibinfo{author}{\bibfnamefont{D.}~\bibnamefont{Kim}},
  \bibinfo{author}{\bibfnamefont{S.}~\bibnamefont{Cho}},
  \bibinfo{author}{\bibfnamefont{N.~P.} \bibnamefont{Butch}},
  \bibinfo{author}{\bibfnamefont{P.}~\bibnamefont{Syers}},
  \bibinfo{author}{\bibfnamefont{K.}~\bibnamefont{Kirshenbaum}},
  \bibinfo{author}{\bibfnamefont{S.}~\bibnamefont{Adam}},
  \bibinfo{author}{\bibfnamefont{J.}~\bibnamefont{Paglione}}, \bibnamefont{and}
  \bibinfo{author}{\bibfnamefont{M.~S.} \bibnamefont{Fuhrer}},
  \bibinfo{journal}{Nat.~Phys., advance online publication,
  DOI:10.1038/NPHYS2286}  (\bibinfo{year}{2012}).

\bibitem[{\citenamefont{Br\"une et~al.}(2011)\citenamefont{Br\"une, Liu, Novik,
  Hankiewicz, Buhmann, Chen, Qi, Shen, Zhang, and Molenkamp}}]{Brune_PRL11}
\bibinfo{author}{\bibfnamefont{C.}~\bibnamefont{Br\"une}},
  \bibinfo{author}{\bibfnamefont{C.~X.} \bibnamefont{Liu}},
  \bibinfo{author}{\bibfnamefont{E.~G.} \bibnamefont{Novik}},
  \bibinfo{author}{\bibfnamefont{E.~M.} \bibnamefont{Hankiewicz}},
  \bibinfo{author}{\bibfnamefont{H.}~\bibnamefont{Buhmann}},
  \bibinfo{author}{\bibfnamefont{Y.~L.} \bibnamefont{Chen}},
  \bibinfo{author}{\bibfnamefont{X.~L.} \bibnamefont{Qi}},
  \bibinfo{author}{\bibfnamefont{Z.~X.} \bibnamefont{Shen}},
  \bibinfo{author}{\bibfnamefont{S.~C.} \bibnamefont{Zhang}}, \bibnamefont{and}
  \bibinfo{author}{\bibfnamefont{L.~W.} \bibnamefont{Molenkamp}},
  \bibinfo{journal}{Phys.\ Rev.\ Lett.} \textbf{\bibinfo{volume}{106}},
  \bibinfo{pages}{126803} (\bibinfo{year}{2011}).

\bibitem[{Nim()}]{Nimtz}
\bibinfo{note}{G. Nimtz and B. Schlicht, in \emph{Narrow-Gap Semiconductors},
  Springer Tracts in Modern Physics Vol.~98, edited by G. H\"ohler
  (Springer-Verlag, Berlin, 1983), p. 1.}

\bibitem[{\citenamefont{Kriechbaum et~al.}(1984)\citenamefont{Kriechbaum,
  Ambrosch, Fantner, Clemens, and Bauer}}]{Kriechbaum_PRB84}
\bibinfo{author}{\bibfnamefont{M.}~\bibnamefont{Kriechbaum}},
  \bibinfo{author}{\bibfnamefont{K.~E.} \bibnamefont{Ambrosch}},
  \bibinfo{author}{\bibfnamefont{E.~J.} \bibnamefont{Fantner}},
  \bibinfo{author}{\bibfnamefont{H.}~\bibnamefont{Clemens}}, \bibnamefont{and}
  \bibinfo{author}{\bibfnamefont{G.}~\bibnamefont{Bauer}},
  \bibinfo{journal}{Phys. Rev. B} \textbf{\bibinfo{volume}{30}},
  \bibinfo{pages}{3394} (\bibinfo{year}{1984}).

\bibitem[{\citenamefont{Grabecki et~al.}(2005)\citenamefont{Grabecki, Wr\'obel,
  Dietl, Janik, Aleszkiewicz, Papis, Kami\'{n}ska, Piotrowska, Springholz, and
  Bauer}}]{Grabecki_PRB05}
\bibinfo{author}{\bibfnamefont{G.}~\bibnamefont{Grabecki}},
  \bibinfo{author}{\bibfnamefont{J.}~\bibnamefont{Wr\'obel}},
  \bibinfo{author}{\bibfnamefont{T.}~\bibnamefont{Dietl}},
  \bibinfo{author}{\bibfnamefont{E.}~\bibnamefont{Janik}},
  \bibinfo{author}{\bibfnamefont{M.}~\bibnamefont{Aleszkiewicz}},
  \bibinfo{author}{\bibfnamefont{E.}~\bibnamefont{Papis}},
  \bibinfo{author}{\bibfnamefont{E.}~\bibnamefont{Kami\'{n}ska}},
  \bibinfo{author}{\bibfnamefont{A.}~\bibnamefont{Piotrowska}},
  \bibinfo{author}{\bibfnamefont{G.}~\bibnamefont{Springholz}},
  \bibnamefont{and} \bibinfo{author}{\bibfnamefont{G.}~\bibnamefont{Bauer}},
  \bibinfo{journal}{Phys.\ Rev.\ B} \textbf{\bibinfo{volume}{72}},
  \bibinfo{pages}{125332} (\bibinfo{year}{2005}).

\bibitem[{\citenamefont{Grabecki et~al.}(2006)\citenamefont{Grabecki,
  Wr{\'o}bel, Dietl, Janik, Aleszkiewicz, Papis, Kaminska, Piotrowska,
  Springholz, and Bauer}}]{Grabecki_PE06}
\bibinfo{author}{\bibfnamefont{G.}~\bibnamefont{Grabecki}},
  \bibinfo{author}{\bibfnamefont{J.}~\bibnamefont{Wr{\'o}bel}},
  \bibinfo{author}{\bibfnamefont{T.}~\bibnamefont{Dietl}},
  \bibinfo{author}{\bibfnamefont{E.}~\bibnamefont{Janik}},
  \bibinfo{author}{\bibfnamefont{M.}~\bibnamefont{Aleszkiewicz}},
  \bibinfo{author}{\bibfnamefont{E.}~\bibnamefont{Papis}},
  \bibinfo{author}{\bibfnamefont{E.}~\bibnamefont{Kaminska}},
  \bibinfo{author}{\bibfnamefont{A.}~\bibnamefont{Piotrowska}},
  \bibinfo{author}{\bibfnamefont{G.}~\bibnamefont{Springholz}},
  \bibnamefont{and} \bibinfo{author}{\bibfnamefont{G.}~\bibnamefont{Bauer}},
  \bibinfo{journal}{Physica E} \textbf{\bibinfo{volume}{34}},
  \bibinfo{pages}{560 } (\bibinfo{year}{2006}).

\bibitem[{\citenamefont{Kolwas et~al.}(2011)\citenamefont{Kolwas, Grabecki,
  Trushkin, Wr{\'o}bel, Aleszkiewicz, Cywi{\'n}ski, Dietl, Springholz, and
  Bauer}}]{Kolwas_arXiv11}
\bibinfo{author}{\bibfnamefont{K.~A.} \bibnamefont{Kolwas}},
  \bibinfo{author}{\bibfnamefont{G.}~\bibnamefont{Grabecki}},
  \bibinfo{author}{\bibfnamefont{S.}~\bibnamefont{Trushkin}},
  \bibinfo{author}{\bibfnamefont{J.}~\bibnamefont{Wr{\'o}bel}},
  \bibinfo{author}{\bibfnamefont{M.}~\bibnamefont{Aleszkiewicz}},
  \bibinfo{author}{\bibfnamefont{{\L}.}~\bibnamefont{Cywi{\'n}ski}},
  \bibinfo{author}{\bibfnamefont{T.}~\bibnamefont{Dietl}},
  \bibinfo{author}{\bibfnamefont{G.}~\bibnamefont{Springholz}},
  \bibnamefont{and} \bibinfo{author}{\bibfnamefont{G.}~\bibnamefont{Bauer}},
  \bibinfo{journal}{arXiv:1111.2433}  (\bibinfo{year}{2011}).

\bibitem[{\citenamefont{Dietl et~al.}(1994)\citenamefont{Dietl, \'{S}liwa,
  Bauer, and Pascher}}]{Dietl_PRB94}
\bibinfo{author}{\bibfnamefont{T.}~\bibnamefont{Dietl}},
  \bibinfo{author}{\bibfnamefont{C.}~\bibnamefont{\'{S}liwa}},
  \bibinfo{author}{\bibfnamefont{G.}~\bibnamefont{Bauer}}, \bibnamefont{and}
  \bibinfo{author}{\bibfnamefont{H.}~\bibnamefont{Pascher}},
  \bibinfo{journal}{Phys.\ Rev.\ B} \textbf{\bibinfo{volume}{49}},
  \bibinfo{pages}{2230} (\bibinfo{year}{1994}).

\bibitem[{\citenamefont{Kepa et~al.}(2003)\citenamefont{Kepa, Springholz,
  Giebultowicz, Goldman, Majkrzak, Kacman, Blinowski, Holl, Krenn, and
  Bauer}}]{Kepa_PRB03}
\bibinfo{author}{\bibfnamefont{H.}~\bibnamefont{Kepa}},
  \bibinfo{author}{\bibfnamefont{G.}~\bibnamefont{Springholz}},
  \bibinfo{author}{\bibfnamefont{T.~M.} \bibnamefont{Giebultowicz}},
  \bibinfo{author}{\bibfnamefont{K.~I.} \bibnamefont{Goldman}},
  \bibinfo{author}{\bibfnamefont{C.~F.} \bibnamefont{Majkrzak}},
  \bibinfo{author}{\bibfnamefont{P.}~\bibnamefont{Kacman}},
  \bibinfo{author}{\bibfnamefont{J.}~\bibnamefont{Blinowski}},
  \bibinfo{author}{\bibfnamefont{S.}~\bibnamefont{Holl}},
  \bibinfo{author}{\bibfnamefont{H.}~\bibnamefont{Krenn}}, \bibnamefont{and}
  \bibinfo{author}{\bibfnamefont{G.}~\bibnamefont{Bauer}},
  \bibinfo{journal}{Phys.\ Rev.\ B} \textbf{\bibinfo{volume}{68}},
  \bibinfo{pages}{024419} (\bibinfo{year}{2003}).

\bibitem[{\citenamefont{Lechner et~al.}(2005)\citenamefont{Lechner, Springholz,
  Sch\"ulli, Stangl, Schwarzl, and Bauer}}]{Lechner_PRL05}
\bibinfo{author}{\bibfnamefont{R.~T.} \bibnamefont{Lechner}},
  \bibinfo{author}{\bibfnamefont{G.}~\bibnamefont{Springholz}},
  \bibinfo{author}{\bibfnamefont{T.~U.} \bibnamefont{Sch\"ulli}},
  \bibinfo{author}{\bibfnamefont{J.}~\bibnamefont{Stangl}},
  \bibinfo{author}{\bibfnamefont{T.}~\bibnamefont{Schwarzl}}, \bibnamefont{and}
  \bibinfo{author}{\bibfnamefont{G.}~\bibnamefont{Bauer}},
  \bibinfo{journal}{Phys.\ Rev.\ Lett.} \textbf{\bibinfo{volume}{94}},
  \bibinfo{pages}{157201} (\bibinfo{year}{2005}).

\bibitem[{\citenamefont{Grabecki et~al.}(2010)\citenamefont{Grabecki, Kolwas,
  Wr{\'o}bel, Kapcia, Pu{\'z}niak, Jakie{\l}a, Aleszkiewicz, Dietl, Springholz,
  and Bauer}}]{Grabecki_JAP10}
\bibinfo{author}{\bibfnamefont{G.}~\bibnamefont{Grabecki}},
  \bibinfo{author}{\bibfnamefont{K.~A.} \bibnamefont{Kolwas}},
  \bibinfo{author}{\bibfnamefont{J.}~\bibnamefont{Wr{\'o}bel}},
  \bibinfo{author}{\bibfnamefont{K.}~\bibnamefont{Kapcia}},
  \bibinfo{author}{\bibfnamefont{R.}~\bibnamefont{Pu{\'z}niak}},
  \bibinfo{author}{\bibfnamefont{R.}~\bibnamefont{Jakie{\l}a}},
  \bibinfo{author}{\bibfnamefont{M.}~\bibnamefont{Aleszkiewicz}},
  \bibinfo{author}{\bibfnamefont{T.}~\bibnamefont{Dietl}},
  \bibinfo{author}{\bibfnamefont{G.}~\bibnamefont{Springholz}},
  \bibnamefont{and} \bibinfo{author}{\bibfnamefont{G.}~\bibnamefont{Bauer}},
  \bibinfo{journal}{J.\ Appl.\ Phys.} \textbf{\bibinfo{volume}{108}},
  \bibinfo{pages}{053714} (\bibinfo{year}{2010}).

\bibitem[{\citenamefont{Murakami and Kuga}(2008)}]{Murakami_PRB08}
\bibinfo{author}{\bibfnamefont{S.}~\bibnamefont{Murakami}} \bibnamefont{and}
  \bibinfo{author}{\bibfnamefont{S.-i.} \bibnamefont{Kuga}},
  \bibinfo{journal}{Phys.\ Rev.\ B} \textbf{\bibinfo{volume}{78}},
  \bibinfo{pages}{165313} (\bibinfo{year}{2008}).

\bibitem[{\citenamefont{Hsieh et~al.}(2012)\citenamefont{Hsieh, Lin, Liu, Duan,
  Bansil, and Fu}}]{Hsieh_arXiv12}
\bibinfo{author}{\bibfnamefont{T.~H.} \bibnamefont{Hsieh}},
  \bibinfo{author}{\bibfnamefont{H.}~\bibnamefont{Lin}},
  \bibinfo{author}{\bibfnamefont{J.}~\bibnamefont{Liu}},
  \bibinfo{author}{\bibfnamefont{W.}~\bibnamefont{Duan}},
  \bibinfo{author}{\bibfnamefont{A.}~\bibnamefont{Bansil}}, \bibnamefont{and}
  \bibinfo{author}{\bibfnamefont{L.}~\bibnamefont{Fu}},
  \bibinfo{journal}{arXiv:1202.1003}  (\bibinfo{year}{2012}).

\bibitem[{\citenamefont{Fu}(2011)}]{Fu_PRL11}
\bibinfo{author}{\bibfnamefont{L.}~\bibnamefont{Fu}}, \bibinfo{journal}{Phys.\
  Rev.\ Lett.} \textbf{\bibinfo{volume}{106}}, \bibinfo{pages}{106802}
  (\bibinfo{year}{2011}).

\bibitem[{\citenamefont{Volkov and Pankratov}(1985)}]{Volkov_JETP85}
\bibinfo{author}{\bibfnamefont{B.~A.} \bibnamefont{Volkov}} \bibnamefont{and}
  \bibinfo{author}{\bibfnamefont{O.~A.} \bibnamefont{Pankratov}},
  \bibinfo{journal}{JETP Lett.} \textbf{\bibinfo{volume}{42}},
  \bibinfo{pages}{178} (\bibinfo{year}{1985}).

\bibitem[{\citenamefont{Korenman and Drew}(1987)}]{Korenman_PRB87}
\bibinfo{author}{\bibfnamefont{V.}~\bibnamefont{Korenman}} \bibnamefont{and}
  \bibinfo{author}{\bibfnamefont{H.~D.} \bibnamefont{Drew}},
  \bibinfo{journal}{Phys.\ Rev.\ B} \textbf{\bibinfo{volume}{35}},
  \bibinfo{pages}{6446} (\bibinfo{year}{1987}).

\bibitem[{\citenamefont{Agassi and Korenman}(1988)}]{Agassi_PRB88}
\bibinfo{author}{\bibfnamefont{D.}~\bibnamefont{Agassi}} \bibnamefont{and}
  \bibinfo{author}{\bibfnamefont{V.}~\bibnamefont{Korenman}},
  \bibinfo{journal}{Phys.\ Rev.\ B} \textbf{\bibinfo{volume}{37}},
  \bibinfo{pages}{10095} (\bibinfo{year}{1988}).

\bibitem[{\citenamefont{Pankratov}(1990)}]{Pankratov_SST90}
\bibinfo{author}{\bibfnamefont{O.~A.} \bibnamefont{Pankratov}},
  \bibinfo{journal}{Semicond. Sci. Technol.} \textbf{\bibinfo{volume}{5}},
  \bibinfo{pages}{S204} (\bibinfo{year}{1990}).

\bibitem[{\citenamefont{Ringel et~al.}(2011)\citenamefont{Ringel, Kraus, and
  Stern}}]{Ringel_arXiv11}
\bibinfo{author}{\bibfnamefont{Z.}~\bibnamefont{Ringel}},
  \bibinfo{author}{\bibfnamefont{Y.~E.} \bibnamefont{Kraus}}, \bibnamefont{and}
  \bibinfo{author}{\bibfnamefont{A.}~\bibnamefont{Stern}},
  \bibinfo{journal}{arXiv:1105.4351}  (\bibinfo{year}{2011}).

\bibitem[{\citenamefont{Mong et~al.}(2012)\citenamefont{Mong, Bardarson, and
  Moore}}]{Mong_PRL12}
\bibinfo{author}{\bibfnamefont{R.~S.~K.} \bibnamefont{Mong}},
  \bibinfo{author}{\bibfnamefont{J.~H.} \bibnamefont{Bardarson}},
  \bibnamefont{and} \bibinfo{author}{\bibfnamefont{J.~E.} \bibnamefont{Moore}},
  \bibinfo{journal}{Phys.\ Rev.\ Lett.} \textbf{\bibinfo{volume}{108}},
  \bibinfo{pages}{076804} (\bibinfo{year}{2012}).

\bibitem[{\citenamefont{Liu et~al.}(2010{\natexlab{a}})\citenamefont{Liu,
  Zhang, Yan, Qi, Frauenheim, Dai, Fang, and Zhang}}]{Liu_crossover_PRB10}
\bibinfo{author}{\bibfnamefont{C.-X.} \bibnamefont{Liu}},
  \bibinfo{author}{\bibfnamefont{H.}~\bibnamefont{Zhang}},
  \bibinfo{author}{\bibfnamefont{B.}~\bibnamefont{Yan}},
  \bibinfo{author}{\bibfnamefont{X.-L.} \bibnamefont{Qi}},
  \bibinfo{author}{\bibfnamefont{T.}~\bibnamefont{Frauenheim}},
  \bibinfo{author}{\bibfnamefont{X.}~\bibnamefont{Dai}},
  \bibinfo{author}{\bibfnamefont{Z.}~\bibnamefont{Fang}}, \bibnamefont{and}
  \bibinfo{author}{\bibfnamefont{S.-C.} \bibnamefont{Zhang}},
  \bibinfo{journal}{Phys.\ Rev.\ B} \textbf{\bibinfo{volume}{81}},
  \bibinfo{pages}{041307} (\bibinfo{year}{2010}{\natexlab{a}}).

\bibitem[{\citenamefont{Linder et~al.}(2009)\citenamefont{Linder, Yokoyama, and
  Sudb\o{}}}]{Linder_PRB09}
\bibinfo{author}{\bibfnamefont{J.}~\bibnamefont{Linder}},
  \bibinfo{author}{\bibfnamefont{T.}~\bibnamefont{Yokoyama}}, \bibnamefont{and}
  \bibinfo{author}{\bibfnamefont{A.}~\bibnamefont{Sudb\o{}}},
  \bibinfo{journal}{Phys.\ Rev.\ B} \textbf{\bibinfo{volume}{80}},
  \bibinfo{pages}{205401} (\bibinfo{year}{2009}).

\bibitem[{\citenamefont{Lu et~al.}(2010)\citenamefont{Lu, Shan, Yao, Niu, and
  Shen}}]{Lu_PRB10}
\bibinfo{author}{\bibfnamefont{H.-Z.} \bibnamefont{Lu}},
  \bibinfo{author}{\bibfnamefont{W.-Y.} \bibnamefont{Shan}},
  \bibinfo{author}{\bibfnamefont{W.}~\bibnamefont{Yao}},
  \bibinfo{author}{\bibfnamefont{Q.}~\bibnamefont{Niu}}, \bibnamefont{and}
  \bibinfo{author}{\bibfnamefont{S.-Q.} \bibnamefont{Shen}},
  \bibinfo{journal}{Phys.\ Rev.\ B} \textbf{\bibinfo{volume}{81}},
  \bibinfo{pages}{115407} (\bibinfo{year}{2010}).

\bibitem[{\citenamefont{Shan et~al.}(2010)\citenamefont{Shan, Lu, and
  Shen}}]{Shan_NJP10}
\bibinfo{author}{\bibfnamefont{W.-Y.} \bibnamefont{Shan}},
  \bibinfo{author}{\bibfnamefont{H.-Z.} \bibnamefont{Lu}}, \bibnamefont{and}
  \bibinfo{author}{\bibfnamefont{S.-Q.} \bibnamefont{Shen}},
  \bibinfo{journal}{New J.~Phys.} \textbf{\bibinfo{volume}{12}},
  \bibinfo{pages}{043048} (\bibinfo{year}{2010}).

\bibitem[{\citenamefont{Hewes et~al.}(1973)\citenamefont{Hewes, Adler, and
  Senturia}}]{Hewes_PRB73}
\bibinfo{author}{\bibfnamefont{C.~R.} \bibnamefont{Hewes}},
  \bibinfo{author}{\bibfnamefont{M.~S.} \bibnamefont{Adler}}, \bibnamefont{and}
  \bibinfo{author}{\bibfnamefont{S.~D.} \bibnamefont{Senturia}},
  \bibinfo{journal}{Phys.\ Rev.\ B} \textbf{\bibinfo{volume}{7}},
  \bibinfo{pages}{5195} (\bibinfo{year}{1973}).

\bibitem[{App()}]{Appold79}
\bibinfo{note}{G.~Appold, R.~Grisar, G.~Bauer, H.~Burkhard, R.~Ebert,
  H.~Pacher, H.G.~H{\:a}fele, Proc.~14th Int. Conf. Phys. Semicond. (ed.~B.L.H.
  Wilson, Institute of Physics, Edinburgh 1979), p.~1101.}

\bibitem[{\citenamefont{He and Grassie}(1985)}]{He_JPF85}
\bibinfo{author}{\bibfnamefont{Y.~S.} \bibnamefont{He}} \bibnamefont{and}
  \bibinfo{author}{\bibfnamefont{A.~D.~C.} \bibnamefont{Grassie}},
  \bibinfo{journal}{J.~Phys.~F: Met.~Phys.} \textbf{\bibinfo{volume}{15}},
  \bibinfo{pages}{263} (\bibinfo{year}{1985}).

\bibitem[{\citenamefont{Liu et~al.}(2010{\natexlab{b}})\citenamefont{Liu, Qi,
  Zhang, Dai, Fang, and Zhang}}]{Liu_model_PRB10}
\bibinfo{author}{\bibfnamefont{C.-X.} \bibnamefont{Liu}},
  \bibinfo{author}{\bibfnamefont{X.-L.} \bibnamefont{Qi}},
  \bibinfo{author}{\bibfnamefont{H.}~\bibnamefont{Zhang}},
  \bibinfo{author}{\bibfnamefont{X.}~\bibnamefont{Dai}},
  \bibinfo{author}{\bibfnamefont{Z.}~\bibnamefont{Fang}}, \bibnamefont{and}
  \bibinfo{author}{\bibfnamefont{S.-C.} \bibnamefont{Zhang}},
  \bibinfo{journal}{Phys.\ Rev.\ B} \textbf{\bibinfo{volume}{82}},
  \bibinfo{pages}{045122} (\bibinfo{year}{2010}{\natexlab{b}}).

\bibitem[{\citenamefont{Dugaev and Petrov}(1994)}]{Dugaev_pssb94}
\bibinfo{author}{\bibfnamefont{V.~K.} \bibnamefont{Dugaev}} \bibnamefont{and}
  \bibinfo{author}{\bibfnamefont{P.~P.} \bibnamefont{Petrov}},
  \bibinfo{journal}{phys.~stat.~sol.~(b)} \textbf{\bibinfo{volume}{184}},
  \bibinfo{pages}{347} (\bibinfo{year}{1994}).

\bibitem[{cos()}]{cosgamma_comment}
\bibinfo{note}{Note that for $\cos 2\gamma \! = \! 0$, i.e.~for $V\! =\! V'$,
  there is no real spin texture. What remains is the pseudospin texture
  described by Eq.~(\ref{eq:H2}), which still provides protection against
  elastic backscattering.}

\bibitem[{\citenamefont{Prokofievaa et~al.}(2010)\citenamefont{Prokofievaa,
  Ravich, Pshenay-Severin, Konstantinov, and
  Shabaldin}}]{Prokofieva_Semiconductors10}
\bibinfo{author}{\bibfnamefont{L.~V.} \bibnamefont{Prokofievaa}},
  \bibinfo{author}{\bibfnamefont{Y.~I.} \bibnamefont{Ravich}},
  \bibinfo{author}{\bibfnamefont{D.~A.} \bibnamefont{Pshenay-Severin}},
  \bibinfo{author}{\bibfnamefont{P.~P.} \bibnamefont{Konstantinov}},
  \bibnamefont{and} \bibinfo{author}{\bibfnamefont{A.~A.}
  \bibnamefont{Shabaldin}}, \bibinfo{journal}{Semiconductors}
  \textbf{\bibinfo{volume}{44}}, \bibinfo{pages}{712} (\bibinfo{year}{2010}).

\bibitem[{\citenamefont{Butch et~al.}(2010)\citenamefont{Butch, Kirshenbaum,
  Syers, Sushkov, Jenkins, Drew, and Paglione}}]{Butch_PRB10}
\bibinfo{author}{\bibfnamefont{N.~P.} \bibnamefont{Butch}},
  \bibinfo{author}{\bibfnamefont{K.}~\bibnamefont{Kirshenbaum}},
  \bibinfo{author}{\bibfnamefont{P.}~\bibnamefont{Syers}},
  \bibinfo{author}{\bibfnamefont{A.~B.} \bibnamefont{Sushkov}},
  \bibinfo{author}{\bibfnamefont{G.~S.} \bibnamefont{Jenkins}},
  \bibinfo{author}{\bibfnamefont{H.~D.} \bibnamefont{Drew}}, \bibnamefont{and}
  \bibinfo{author}{\bibfnamefont{J.}~\bibnamefont{Paglione}},
  \bibinfo{journal}{Phys. Rev. B} \textbf{\bibinfo{volume}{81}},
  \bibinfo{pages}{241301} (\bibinfo{year}{2010}).

\bibitem[{\citenamefont{Khokhlov et~al.}(2000)\citenamefont{Khokhlov, Ivanchik,
  Raines, Watson, and Pipher}}]{Khokhlov_APL00}
\bibinfo{author}{\bibfnamefont{D.~R.} \bibnamefont{Khokhlov}},
  \bibinfo{author}{\bibfnamefont{I.~I.} \bibnamefont{Ivanchik}},
  \bibinfo{author}{\bibfnamefont{S.~N.} \bibnamefont{Raines}},
  \bibinfo{author}{\bibfnamefont{D.~M.} \bibnamefont{Watson}},
  \bibnamefont{and} \bibinfo{author}{\bibfnamefont{J.~L.}
  \bibnamefont{Pipher}}, \bibinfo{journal}{Appl.\ Phys.\ Lett.}
  \textbf{\bibinfo{volume}{76}}, \bibinfo{pages}{2835} (\bibinfo{year}{2000}).

\end{thebibliography}
\end{document}